\documentclass{edbk} % Computer Modern font calls
\usepackage{epsfig}

\begin{document}

%\draft

\turnoff
\setcounter{secnumdepth}{2}
\setcounter{tocdepth}{2}
\normallatexbib

\articletitle[Clustering from N/Z=1 to N/Z=2]
{Clustering in nuclei \\
 from N/Z=1 to N/Z=2 \footnote{
Invited talk presented at ``New Physics 2000'' conference
{\bf The Nucleus: New Physics for the New Millenium}
National Accelerator Centre, Faure, Cape Town, South Africa
18--22 January 1999
in {\em The Nucleus: New Physics for the New Millenium}, Proc.
Int. Conf. held at Faure, Cape Town, January 1999, ed. F.D. Smit,
R. Lindsay and S.V. F\"{o}rtsch, ISBN 0-306-46302-4, (Kluwer
Academic/Plenum, New York, 1999)
 }}

\author{Wilton N. Catford}

%% affil, email, and abstract are optional
\affil{Department of Physics\\
University of Surrey\\
Guildford GU2~5XH, UK}
%\email{w.catford@surrey.ac.uk}

%% optional, to supply a shorter version of the title for the running head:
%%\chaptitlerunninghead{Clustering from N/Z=1 to N/Z=2}

\begin{abstract}
The clustering of nucleons in nuclei is a widespread but elusive phenomenon for
study. Here, we wish to highlight the variety of theoretical approaches,
and demonstrate how they are mutually supportive and complementary.
On the experimental side, we  describe recent advances in the study
of the classic cluster nucleus $^{24}$Mg. Also, recent studies of clustering
in nuclei approaching the neutron drip line are described.
In the region near N/Z=2, both theory and experiment now suggest that
multi-centre cluster structure is important, in particular for
the very neutron rich beryllium isotopes.
\end{abstract}

\section[Types of clustering]{Introduction: types of clustering}
Rarely has a topic in nuclear physics attracted as much misunderstanding
and debate as clustering behaviour in nuclei. Nuclear clusters are like
Schr\"{o}dinger's cats, being both in and out of existence until they are
observed, and in some ways it is even more difficult to untangle their true
behaviour from the process of observation. In some instances, such as the
spectroscopic factors derived from $\alpha $-particle transfer reactions,
the quantitative data give a sure signature of
clustering \cite{Arima-Treatise}. More often,
the signature is less definite but the cluster model leads to a consistent
and intuitive understanding of the data. Taken as a whole, the evidence
is compelling that clustering effects are important in nuclei across
the periodic table, and particularly amongst the lighter nuclei below calcium.
The richness of the subject has ensured a steady flow of recent review articles
\cite{Horiuchi,FreerMerchant,Wuosmaa,Langanke}.

The existence of $\alpha $-particle emission from nuclei suggests that
pre-formed $\alpha $-particles might exist inside them.
From very early on, however,
it has been known \cite{Wheeler,Wefelmeier}
that it is overly naive to suppose that any nucleus is
composed of real $\alpha $-particles bound together. In a nucleus such as
$^{16}$O, say, the $\alpha $-particles would overlap and
effectively tear each other
apart through the strong interaction. Still, as we shall see,
the tendency towards clustering persists and it is driven by the tight
binding of the maximally symmetric system of 2 protons and 2 neutrons
coupled to L\,=\,S\,=\,0\,; the $(0s_{1/2})^{4}$ shell model configuration.
The Ikeda diagram \cite{Ikeda-Takigawa-Horiuchi} predicts widespread clustering
structure, and in particular $\alpha $-cluster structure,
for states in nuclei at energies near the various cluster separation thresholds.

If we take $^{12}$C as an example, it is appealing to imagine three
interacting $\alpha $--clusters dragging past each other in orbital motion,
and exchanging nucleons through their strong interaction.
For a given nucleon, its wavefunction is not confined to a single
$\alpha $--cluster but is shared between different clusters. Perhaps it is
reasonable to think of this as a kind of Bose-Einstein condensate of
$\alpha $-particles. The overall A-body wavefunction must be antisymmetric,
and it is the action of the
antisymmetrization operator that effectively shares the
wavefunction of each nucleon between the cluster mass centres. The average
particle density distribution for the properly antisymmetrized intrinsic
wavefunction of the nucleus continues to show the cluster-like appearance.

In some theoretical calculations, it is convenient to assume the
existence of cluster centres {\em a priori}, and subsequently to deal
with the antisymmetrization either exactly or according to
an approximate prescription. Of course, the clusters need not in general
be simply $\alpha $-like. Further, models have recently been developed that
allow the general unconstrained A-body problem to be solved using a
variational method, wherein the individual nucleons are allowed to move
independently with a two-body force acting between them and are not
forced into clusters at all. The
results of such calculations show a natural tendency
towards clustering behaviour in certain cases and mean field
behaviour in others. One exciting prediction is that the light neutron
rich nuclei such as beryllium, boron and carbon exhibit multicentre
cluster structure as neutrons are added out towards the drip-line.
This clustering may be a significant factor in determining the
structure of neutron haloes in nuclei such as $^{11}$Be and $^{14}$Be.
In any case, the models of halo nuclei which assume a core
plus one or two valence nucleons are examples of cluster models, where
the core is treated as a cluster.

Clustering behaviour has been seen in a wide variety of guises
\cite{ShelineWildermuth,CsehScheid} across the
nuclear chart, cf. fig.1, including:
\begin{itemize}
\item {\bf Very light nuclei:} in cases such as $^8$Be ($\alpha + \alpha$)
\cite{Wheeler-Be8} and
$^7$Li ($\alpha + t$) \cite{Buck-li7}, or even $^{12}$C ($3\alpha$) \cite{Brink},
there are small subsystems of
the nucleus that are quite tightly bound and there is a tendency for this
structure to develop and to be competitive in energy with a  mean field
configuration. Additional neutrons can be added into molecular orbitals in $^8$Be
using the model of Abe \cite{Abe} or otherwise
and account for the properties of $^9$Be and $^{10}$Be \cite{Seya,vonO-1},
\item {\bf Magic core plus orbiting cluster:} in nuclei just beyond strong shell
closures, the valence nucleons can behave like a cluster orbiting outside of the
closed shell nucleus, giving characteristic rotational spectra, which have
been observed near the $^{16}$O, $^{40}$Ca, $^{90}$Zn and $^{208}$Pb doubly magic
closures \cite{buck-closed-shell-1}; this is discussed
in the next section,
\item {\bf Normal core plus nucleon halo:} when for example the final neutron
is extremely weakly bound within the mean field of the nucleus, then its
wavefunction naturally tends to extend substantially beyond the edges of the
nuclear potential into the classically forbidden region, particularly if it has
a low angular momentum \cite{Karsten}; an even more interesting case is with two
nucleons in the halo, such as in Borromean systems \cite{Borromean},
\item {\bf Large-scale clustering in low-lying levels:} complex nuclei have
been successfully modelled with the assumption that they comprise two mutually
orbiting nuclei of comparable masses, with the predicted spectra showing
remarkable agreement from the ground state all the way up to the separation
energy \cite{Kabir,Buck-mg24,BaldockBuck};
this, despite the fact that the components must surely overlap,
\item {\bf Large-scale clustering at high excitation:} resonances seen at
energies near the Coulomb barrier in heavy ion scattering
\cite{Bromley,Treatise,Cindro},
particularly between
$\alpha $-conjugate nuclei, seem to indicate molecular-like states in which the
component nuclei orbit at a distance where they just graze against each other
\cite{Cindro,Feshbach}; this can also be studied in breakup reactions,
\item {\bf Complete alpha-particle condensation:} an extreme form of clustering
behaviour is predicted in some models for $A=4n$ nuclei, wherein the entire
nucleus behaves not as one liquid drop but as a condensation into $n$ separate
$\alpha $-like droplets \cite{Brink,Rae-International},
and when extra excitation energy is added to these nuclei
the configuration of the mass centres can unwrap and become spatially extended;
the limiting case is represented by linear chain states, which have been
predicted over a wide range of masses \cite{Merchant-chains}.

\end{itemize}

\begin{figure}[h]
\vskip2pt
\caption{A wide variety of different types of clustering behaviour have
been identified in nuclei, from small clusters outside of a closed shell,
to complete condensation into $\alpha $-particles, to halo nucleons outside
of a normal core.}
%\mockfigure{0.45\textheight}{overhead from powerpoint}
\begin{center}
~\psfig{file=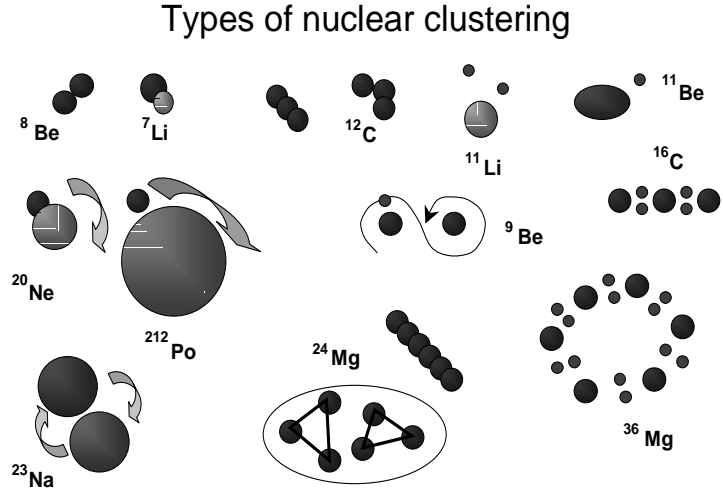,height=0.35\textheight}
\end{center}
\end{figure}

The various different models that have been developed are described
in the reviews \cite{Horiuchi,FreerMerchant,Wuosmaa,Langanke}.
One theoretical approach is to treat the clusters within the nucleus as
real clusters and to solve the many-body scattering problem, for example in the
three-body case by solving the three-body Fadeev equations \cite{Borromean}.
This can be
successful provided that the clusters  avoid overlapping.
Alternatively, the cluster structure can be assumed but the exchange of
nucleons can be allowed, and the wavefunctions properly
antisymmetrized. Such models
using the resonating group method (RGM) \cite{Langanke,Wildermuth} and
the generator coordinate method (GCM) \cite{Descouvemont} have been extensively
applied to light nuclei and in particular for calculations of astrophysically
important processes. Group theoretic approaches have been used to analyse the
symmetry properties of cluster states \cite{Cseh}. Approximate two-body models
have been applied to nuclei where a cluster orbits a core \cite{BuckDoverVary},
with the orbital quantum numbers chosen in such a way as to approximately
satisfy the Exclusion Principle by avoiding the overlap of the core and cluster
wavefunctions. Another model that has been widely applied is the Brink-Bloch
$\alpha $-cluster model \cite{Brink,Rae-International}, in which nucleons are
assumed to be bound in harmonic potential wells forming
$\alpha $-like clusters with L=S=0. An
effective two-body force acts between the clusters and a fully antisymmetrized
A-body Slater determinant wavefunction is derived.
The equilibrium positions of the mass centres
are determined in a variational calculation.
Time dependence can be introduced to this model to give the time-dependent
cluster model (TDCM) \cite{Bauhoff}. Taking this one step further, individual
unconstrained nucleons can be treated in a similar fashion with a two-body
force acting between independently moving nucleons.
This approach has been developed into models known as Fermionic molecular
dynamics \cite{Feldmeier} and antisymmetrized molecular dynamics (AMD)
\cite{Kanada-dissolution}. The quantities that can be calculated in the various
models vary from case to case, but energy levels,
electromagnetic moments and transition densities
have all been very successfully predicted.
An exciting recent development in theory, via the
AMD model, is the ability to predict systematically the persistence of
clustering in systems
that differ from the simple $A=4n, T_z = 0$ $\alpha $-like nuclei, and this
will be discussed below.

\section[Key results from the models]
{Review of some key results from the models}

\subsection{Positive and negative parity rotational bands}
%\paragraph{This is the paragraph}

\begin{figure}[h]
\vskip2pt
\caption{\label{ne-ti}The energy offset between  positive and negative parity
states in rotational bands, shown here for the ground state bands in
$^{20}$Ne and $^{44}$Ti, supports the cluster model.
%Adapted from Buck, Johnston, Merchant and Perez, Phys. Rev. C52 (1995) 1840.
}
% \protect\cite{buck-closed-shell}.}
%\mockfigure{0.4\textheight}{adapted spectra from Buck via xfig}
~\\~\psfig{file=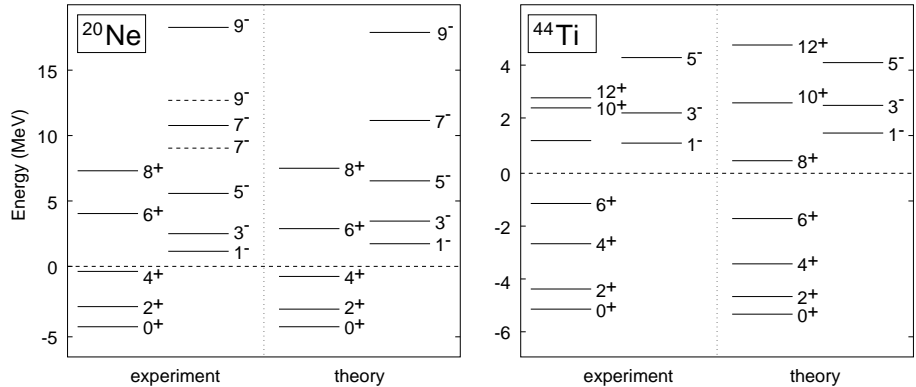,width=\textwidth}
\end{figure}

In $^{20}$Ne for example, there is a negative parity K=0 rotational band with
the same moment of inertia as the positive parity ground state band, but
displaced upwards in energy. The natural explanation for this in terms of the
cluster model was first put forward by Horiuchi and Ikeda \cite{Horiuchi-Ikeda}.
These bands correspond to an $\alpha $-particle orbiting an $^{16}$O
core \cite{WildermuthKenellopoulos}. Assuming
that it is possible to start with an $\alpha $-particle
on one side of the core and to pass continuously,
via the radial motion of the cluster, to the
situation with it on the opposite side of the core, then this implies
a symmetric
potential energy function that has a barrier at zero separation. The intrinsic
solutions centred on the two minima must be properly symmetrized to obtain
states of good parity, and the resultant states
of different parity have different behaviour
in the region of the central barrier. The negative parity state vanishes
at zero but the positive parity state does not.
The net effect is to shift negative
parity states upwards in energy compared to the positive parity states. This
behaviour is clearly seen in $^{20}$Ne and also in $^{44}$Ti, where an
$\alpha $-particle orbits a $^{40}$Ca core
(see fig. \ref{ne-ti}, \cite{buck-closed-shell-2}).
In fact, if the barrier between the two reflection symmetric states (before
parity projection) became infinitely large, then the probability of being at
zero would be zero for the positive parity solution also,
and there would be no shift
between the different parity rotational states in that case.
This closely parallels the situation for octupole deformation in nuclei,
where intrinsic octupole
states can be continuously transformed into their degenerate mirror images.
In their review of intrinsic reflection asymmetry, Butler and
Nazarewicz \cite{ButlerWitek} point out that the $sd$-shell region provided
the first evidence for reflection asymmetry in rotating nuclei, and they
include a detailed bibliography for studies of $^{16}$O, $^{18}$O,
$^{20}$Ne, $^{24}$Mg, $^{28}$Si, $^{32}$S, $^{40}$Ca and $^{44}$Ti,
concentrating on the octupole aspects.

Another important point to note is that
the level scheme of {\em core\,+\,cluster} nuclei
is not the only property which is well reproduced by
cluster model calculations.
Using models such as that of Buck, Dover and Vary \cite{BuckDoverVary},
electromagnetic transition densities can be calculated correctly, without the
need for effective charges such as required in shell model calculations.
These calculations have been extended successfully to include $\alpha$-particle
states above the $^{90}$Zr and $^{208}$Pb double shell closures
\cite{buck-closed-shell-1}.
Alpha-decay widths and low energy elastic scattering cross sections have been
reproduced in a unified treatment for $^{20}$Ne and
$^{44}$Ti \cite{buck-closed-shell-2}.
The model can be extended successfully to nuclei in which exotic
clusters (those which have been observed in radioactivity, such as $^{24}$Ne
\cite{RoseJones}\,) orbit outside of a $^{208}$Pb core. Buck {\em et al}
successfully reproduced the systematics of exotic decay half-lives
\cite{Buck-exotic-halflives} and also the excitation energies for rotational
bands (including Coriolis antistretching) and electromagnetic transition strengths for
collective quadrupole and octupole transitions \cite{buck-actinides} across 19
even-even actinide nuclei.

\subsection{The structure of \,$^{12}$C}

The nucleus $^{12}$C has long been treated theoretically as a triangular
arrangement of $\alpha$-like structures \cite{Brink}. Models
that presuppose the existence of clusters, such as that of Brink \cite{Brink},
have recently received extra theoretical support from two different directions.
In the first instance, the quantum mechanical three-body problem, including the
Coulomb force, was solved for real $\alpha $-particles using a hyperspherical
expansion of the Fadeev equations \cite{Fedorov}. A phenomenological three-body
force was included to account for Pauli and polarization effects. With
parameters chosen to reproduce the energy of the
three-body $(0_2^+,7.65{\rm ~MeV})$
scattering state, the width of the $0_2^+$ and the energy of the bound $0_1^+$
ground state were reasonably reproduced. The
calculations predicted the most likely relative spatial configurations of the
three $\alpha $-particles in the scattering problem. The ground state was an
almost pure equilateral triangle with sides of 3.0 fm, whereas the excited
(unbound) state was a superposition of several configurations with the main one
being a somewhat larger and flattened triangle.
% with sides of 3.9, 4.2 and 6.1 fm.
The ground state
result is quite close to the geometry found in a GCM calculation which used an
equilateral triangle with the length of the sides being a fitted parameter
\cite{DufourDescouvemont}. The
$(0_2^+,7.65{\rm ~MeV})$ state was once thought \cite{Morinaga-766chain} to be an example
of a linear $\alpha $-particle chain state of the type postulated by Morinaga
\cite{Morinaga}.
Later work showed that the $0_2^+ - 2_2^+$ energy gap was not consistent with a
linear 3$\alpha $ linear chain structure \cite{Fujiwara} and a bent
chain  was proposed \cite{Horiuchi-Ikeda-Suzuki}. The three-body
Fadeev calculations support this general interpretation although they
imply a hinge angle of around $100^\circ $ in the dominant
configuration rather than the $150^\circ $
favoured by earlier calculations \cite{Friedrich}.

Another recent theoretical treatment of $^{12}$C is that of Kanada-En'yo
\cite{Kanada-carbon12}. She used the antisymmetrized molecular dynamics (AMD)
model \cite{Horiuchi,Kanada-dissolution}, extended to perform the
variational part of the calculation after projecting states of good parity
and angular momentum. In these calculations, the relative geometry of the
individual nucleons was not constrained at all {\em a priori}, i.e no
clustering was imposed on the nucleus.
The predicted spectrum of levels and their transition strengths
show several advantages over  cluster calculations, in particular
for the relative properties of the $0_1^+$ ground state and the $0_2^+$ state.
An analysis of the wavefunctions obtained  \cite{Kanada-carbon12}
shows that the $\alpha $-clusters are significantly dissociated
in the ground state, but the results also support
a well-developed 3$\alpha $ structure for other rotational bands.

\subsection{Dissolution of alpha particles}

As discussed above,  $^{20}$Ne displays a rotational band spectrum that
indicates an $^{16}$O+$\alpha $ structure. Recently, within the framework of
the AMD model, it has become possible to study theoretically the degree of
clustering in the
different states of the band \cite{Kanada-dissolution}. These new calculations
support many of the earlier deductions \cite{Arima-advances} concerning
reduced clustering in the ground state band towards its termination at spin
$8^+$. A cranking method was used to study states along the yrast line in
$^{20}$Ne, starting from the overall minimum energy solution for the $0^+$
ground state. Nucleon density distributions were then calculated for the
rotating intrinsic states before parity projection. For the lower states in the
band, the structure was found to be prolate and approximately axially symmetric,
consistent with an $^{16}$O+$\alpha $ structure. Furthermore, the $^{16}$O part
of the mass density comprised a tetrahedral arrangement of four $\alpha $-like
mass centres arranged as predicted by conventional cluster models of $^{16}$O.
The negative parity states showed particularly well-developed clustering.
For spins of $7^-$ and above, the $^{16}$O+$\alpha $ structure tended to
dissolve and other structures including oblate shapes and
$^{12}$C$\alpha \alpha$
cluster configurations began to appear. Although they have certain
deficiencies, for example the negative parity energies are not well reproduced,
the calculations give a fascinating insight into the evolution of clustering
with excitation energy and spin. At the higher excitation energies they help to
delineate the limits of applicability for cluster models.
At the lower energies the AMD results strongly support the validity of
assuming a cluster structure, and hence they support the
predictions of other quantities that can be made most effectively using the
cluster models.

\subsection{Large-scale or di-nuclear clustering}

Calculations using an $\alpha $-cluster model for a large nucleus sometimes
give solutions that show a tendency for the nucleus
to be grouped into two or more substantial substructures.
This generalises the behaviour seen for
$^{20}$Ne in the AMD. Indeed, the calculations of Marsh and Rae \cite{MarshRae}
using the Brink model \cite{Brink} show that the ground state of  $^{24}$Mg can
be viewed as two $^{12}$C nuclei in juxtaposition.
In a certain sense, therefore, it is  not too
surprising that the low-lying levels of nuclei such as $^{24}$Mg can be
modelled as two interacting $^{12}$C nuclei. What is truly remarkable, however,
is that such a model \cite{Buck-mg24,BaldockBuck} succeeds with the component
nuclei taking on their free-space properties, even though they presumably
overlap considerably.

In the approach of Buck {\em et al} \cite{Buck-mg24}, $^{24}$Mg is modelled as
a bound state of two $^{12}$C nuclei which can be internally
excited.  The internal energy levels and the
electromagnetic transition strengths between
them are taken to be those for real, free $^{12}$C nuclei. Coupled channels
calculations are performed to deduce the level scheme up to a maximum
energy given by
the sum of the Coulomb barrier and the $^{12}$C+$^{12}$C separation energy.
Almost the entire experimentally known T=0 spectrum is reproduced, both for
positive and negative parity states. The calculated electromagnetic strengths for
quadrupole transitions between the low-lying positive parity states are also
in good agreement with experimental values, without the need to introduce
effective charges. The model has been extended to describe $^{23}$Na in terms of
$^{12}$C and $^{11}$B, each with their free-space properties \cite{Kabir}.
Excellent agreement is achieved with experimental E2 and M1 transition
strengths, and the model has also been extended successfully to predict
polarization observables \cite{KabirJohnson}. Whether such models
have a physical significance beyond their computational convenience
is not clear.

\subsection{Natural clustering tendency in neutron rich systems}

When additional neutrons are added to the $^8$Be core of two $\alpha$-clusters,
molecular states in $^9$Be and $^{10}$Be are formed
\cite{Seya,vonO-1,Okabe}. Similarly, protons can be added and account
for structures in the nuclei $^9$B and $^{10}$B \cite{vonO-1,Okabe-2}. These
approaches can be extended to the more neutron rich isotopes of
beryllium and boron
\cite{Seya,vonO-2,vonO-3} and to analogous 3$\alpha $ systems in the carbon isotopes
\cite{vonO-2}. Thus, Nature edges towards the speculative suggestion of
Wilkinson \cite{Wilkinson} that complete rings of $\alpha $-clusters might
exist, like necklaces, held together by neutron pairs in covalent bonds
between the clusters (cf. fig.1).
For now, however, necklace states remain pure speculation
and even simple chain states of $\alpha $-clusters (which are predicted to
exist at high excitations over a range of nuclei \cite{Merchant-chains}\,) are
not compellingly in evidence experimentally beyond $^8$Be.

Von Oertzen \cite{vonO-1,vonO-2} has estimated the excitation energies of molecular
chain states in neutron rich beryllium, boron, carbon  and oxygen,
where the $\alpha $-cluster centres are bound together with covalent bonds in
analogy with the well-known \cite{Seya,Okabe} $\alpha {\rm :2n:}\alpha $ states
near 6 MeV in $^{10}$Be. The
chain states are expected to be at higher excitation energies in the more
exotic nuclei \cite{vonO-1,vonO-2},
although other clustering (e.g. $^{12}$Be=$^6$He+$^6$He) could
be anticipated at lower excitations.

But what of the wider significance of clustering in the
lowest energy levels for
light, neutron rich nuclei? An exciting development in the study of these nuclei
has been the AMD predictions for ground state nucleon densities. Calculations
have been performed for isotopic chains extending to the neutron dripline for
ground states of
lithium and beryllium \cite{Kanada-Li-Be,Dote-Be}, boron \cite{Kanada-boron}
and carbon \cite{Kanada-C-nrich,Kanada-C-moments} and
have been summarised by Horiuchi \cite{Horiuchi}. As might be expected,
a tendency towards clustering is predicted near the N=Z line and a
tendency towards more mean-field behaviour near the magic number N=8. However,
for N$>$8 the clustering appears to return and in some cases to be enhanced.
For example, in boron isotopes the proton intrinsic matter distributions show
marked and increasing Li-He clustering in the isotopes $^{15,17,19}$B. Across
isotopic chains, the AMD calculations have been tested against
measured ground state
properties and they are found to reproduce well the radii, binding energies and
magnetic moments, except in the case of known halo states \cite{Horiuchi}.
Similarly, the excited state energy spectra and the electromagnetic transition
strengths between levels are reproduced quite well, except for halo levels.

The structure of excited states, and the evolution of shape within a given
nucleus have been studied in more detail in very recent work with the AMD model
for $^{10}$Be \cite{Kanada-Puri} and $^{12}$C \cite{Kanada-carbon12}.
The AMD calculations highlight very clearly the differences in structure, for
example between the ground state of $^{10}$Be and the quartet of molecular
states near 6 MeV \cite{Kanada-Puri}. They allow the coexistence of cluster
structure and mean field structure to be investigated.
Otherwise, in the explicitly
molecular models  \cite{vonO-2,vonO-3} the relationship
between the lowest states
and the excited molecular states requires careful interpretation.

\section{Recent experimental results}

\subsection{Breakup of \,$^{24}$Mg to $^{12}$C+$^{12}$C}

The breakup of excited $^{24}$Mg, $^{28}$Si and $^{32}$S nuclei into $^{12}$C
and $^{16}$O fragments, following nuclear excitation processes, has been studied
in detail for more than a decade \cite{FreerMerchant}. The widths of the $^{24}$Mg
breakup states
are of order 100--150 keV \cite{high-res} which is comparable to the expected
statistical width \cite{Shapira}. On the other hand
the breakup states and the resonances populated in $^{12}$C + $^{12}$C
scattering near the barrier appear to be closely linked \cite{high-res},
and the scattering resonances have been shown to have
enhanced partial widths for $^{12}$C decay.
The scattering resonances are also correlated between different reaction
channels and they $\gamma $-decay
to the ground state of $^{24}$Mg \cite{Treatise}.

\begin{figure}[h]
\vskip2pt
\caption{\label{martin}Montage of excitation spectra for breakup states seen in
$^{24}$Mg depopulated by
(a) $^{16}$O$^8$Be  or
(b) $^{12}$C$^{12}$C decay.
Also shown, statistical partial width predictions
as a function of excitation energy and spin, for (c) $^{12}$C$^{12}$C
relative to $^{16}$O$^8$Be   and
(d) for $^{12}$C$^{12}$C as a fraction of the total width.}
%\mockfigure{0.7\textheight}{composite figure from Freer paper}
~\\
\begin{center}
~\psfig{file=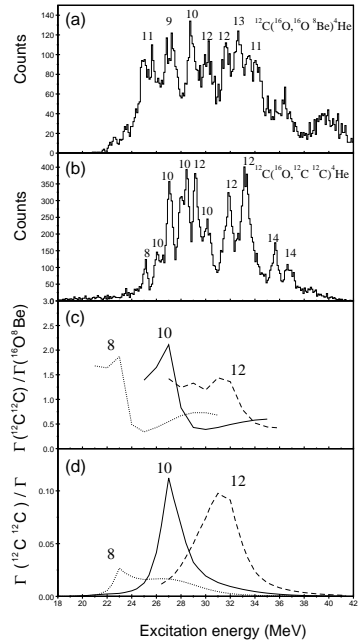,height=0.45\textheight}
\end{center}
\end{figure}

For the breakup states, there has never been a direct partial width
measurement. Recently, Freer {\em et al.} have attempted to obtain such
information indirectly, by comparing the probabilities for the decay of
$^{24}$Mg states into the $^{12}$C+$^{12}$C and $^{16}$O+$^8$Be channels
\cite{Freer}. States in $^{24}$Mg were populated in the reaction
$^{12}$C($^{16}$O,$\alpha $)$^{24}$Mg at several beam energies and the spins
were measured using the angular correlations of the sequential breakup. The
energy-spin systematics followed the grazing angular momentum for two $^{12}$C
nuclei (as expected for molecular states). However, it was also noted that
the statistical decay probabilities for fission to two ground state $^{12}$C
nuclei follow
similar systematics, peaking as the barrier penetration probabilities balance
against the opening of competing channels (cf. Fig.\ref{martin}).
In a quantitative analysis, it
appeared that some states decayed to the two fission channels with the statistical
ratio of probabilities, but that in addition there were several states that
favoured the $^{12}$C+$^{12}$C channel. The situation may
have been particularly complicated
because the breakup states were populated using the ($^{16}$O,$\alpha $)
reaction, which has a compound nuclear mechanism according to its angular distribution
\cite{Freer-oxy-carbon-1}. The mechanism in $^{12}$C($^{24}$Mg,$^{24}$Mg$^*$)$^{12}$C
or in transfer reactions may be more selective of true molecular states.
Clearly, there is a lot more work still to be done in the measurement of
partial decay widths for breakup states.

\subsection{Resonances in mutual $3^-$ scattering of \,$^{12}$C+$^{12}$C}

\begin{figure}[h]
\vskip2pt
\caption{\label{steve} Resonances seen in $^{12}$C+$^{12}$C scattering to
the mutual $3^-_1$ channel: (a) the angular correlations indicate fully aligned
angular momentum in the final state as shown schematically, (b) the energy-spin
systematics are consistent with $\alpha $-cluster predictions for a very
deformed triaxial state in $^{24}$Mg.}
%\mockfigure{0.4\textheight}{3-/3- rotational bands and orbiting triangles}
~\\~\psfig{file=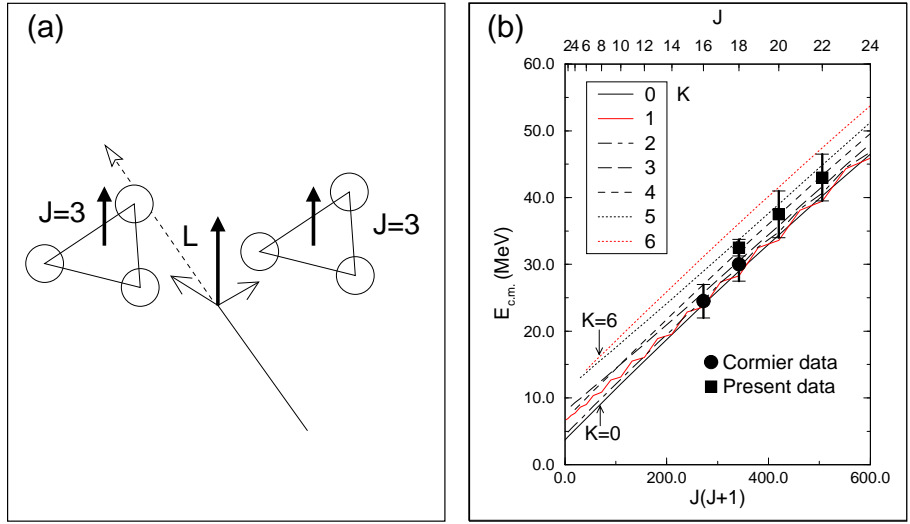,width=\textwidth}
\end{figure}

Chappell {\em et al.} \cite{Chappell}
have recently measured excitation functions and angular
correlations for $^{12}$C+$^{12}$C scattering leading to mutual excitation of
the $^{12}$C$^*$(9.63 MeV; 3$_1^-$) state. In the Brink-Bloch model, both the
ground state and the 3$_1^-$ state have the same intrinsic triangular shape,
and the 3$_1^-$ state has three units of angular momentum around its three-fold
symmetry axis. The new mutual 3$_1^-$ data complement earlier measurements
of the mutual 0$_2^+$ (7.65 MeV) channel \cite{6a-chain} and the 0$_2^+$3$_1^-$
channel \cite{Chappell-early,Wuosmaa-recent} and extend over the energy range
$E_{{\rm cm}}$ = 30--45 MeV.

The 3$_1^-$ state decays by $\alpha $-emission to produce a jet of three
$\alpha $-particles, and these must be reconstructed kinematically in the
analysis to flag 3$_1^-$ production. In fact, the experiment needed to measure
five-fold coincidences and reconstruct the sixth $\alpha $-particle prior to
the reconstructions of the triple jets. In the excitation function, prominent
resonances were observed at $E_{{\rm cm}}$ = 32.5, 37.5 and 43.0 MeV. The
lowest of these corresponds to the energy of the famous resonance in 0$_2^+$0$_2^+$
scattering \cite{6a-chain} that was once believed to
correspond to a shape eigenstate resonance in the form of a 6$\alpha $ linear
chain \cite{shape-eigenstate}, but the 3$_1^-$3$_1^-$ channel is found to be
much stronger \cite{Chappell-early}. In view of the triangular structure of the
3$_1^-$ state, the 6$\alpha $ chain hypothesis is untenable. It can be supposed
that the strength in the 0$_2^+$0$_2^+$ channel arises through the hinge angle
in the 3$\alpha $ 0$_2^+$ state (see earlier), which gives it a partially
triangular structure.

The angular correlation analysis for the 3$_1^-$3$_1^-$ scattering reveals that
the final state is fully aligned in angular momentum, as shown schematically in
fig. \ref{steve}(a). The dominant spins for the three resonances appear to be
18, 20 and 22 respectively. These high spins cannot be supported by $^{24}$Mg
in its normal deformation \cite{Chappell}, so the absorptive scattering
potential is in principal weaker (since only direct reactions and highly deformed
resonances contribute).

Chappell {\em et al.} discuss their results in the context of the Band Crossing
Model \cite{Kondo} using reaction theory, and they use
the Brink-Bloch $\alpha $-cluster
model to give a structural interpretation. The Brink model
predicts \cite{MarshRae} a strongly deformed triaxial state with
$^{12}$C$^{12}$C structure, which corresponds to that labelled F1 by Flocard
{\em et al.} \cite{Flocard}. When this rotates, it gives rise to a number of
bands with different K-values which are close in energy but have very little
K-mixing. It was speculated \cite{Chappell} that the 3$_1^-$3$_1^-$ resonances
(see fig. \ref{steve}(b)) may be associated with the K=6 component, and that
they extend the sequence of broad resonances first observed by Cormier {\em et
al.} \cite{Cormier} in single and mutual 2$_1^+$ scattering.

\subsection{Search for gamma-ray decay of breakup states in $^{24}$Mg}

The $^{12}$C+$^{12}$C states seen in breakup and resonance reactions appear to
show rotational $J(J+1)$ systematics. This could be a genuine structural effect
or it could be due to some kinematic selection. To distinguish these
possibilities, the observation of decay $\gamma $-rays connecting the states
would be useful, but this is exceptionally difficult. One experiment
has so far succeeded in obtaining results \cite{Elanique}. A well known $10^+$
resonance at $E_{{\rm cm}}=16.45$ MeV was populated in $^{12}$C+$^{12}$C
scattering. Particles were detected in coincidence using two position sensitive
detectors to record their energy and angle. The complete binary kinematics
allowed the masses of the particles in the exit channel to be determined, and
the $^{12}$C+$^{12}$C channel could be selected. An array of 74 BaF$_2$
detectors (Ch\^ateau de Cristal) recorded coincident $\gamma $-rays. The
reaction Q-value was reconstructed and the missing kinetic energy was compared
with the $\gamma $-ray energy. If the resonance $\gamma $-decayed to a lower
state in $^{24}$Mg, that subsequently decayed to
$^{12}$C$_{{\rm gs}}$+$^{12}$C$_{{\rm gs}}$, then the compared
energies should be equal. The principal
complication and source of background comes from events in which the breakup
precedes the $\gamma $-decay, that is when the resonance decays to one (or even
two) excited $^{12}$C fragments. This is dominated by the 4.43 MeV, $2_1^+$ channel.
No simple discrimination of these events, other than by $\gamma $-ray energy,
can be made. The result from the experiment was quoted as
$\Gamma _\gamma /\Gamma = 1.2 \times 10^{-5}$ which, with certain assumptions,
corresponds to a transition strength of order 100 W.u.  These numbers are
comparable to the predictions for a molecular band
\cite{Elanique,ChandraMosel}, but because of the low statistics and the
background uncertainties it seems most appropriate
to interpret them as upper limits.
More experiments are definitely required.

\subsection{Two-centre effects in neutron-rich Be isotopes}

\begin{figure}[h]
\vskip2pt
\caption{\label{be12-fig} Inelastic scattering of a radioactive beam of
$^{12}$Be has shown breakup of $^{12}$Be into $^6$He$^6$He, and the spins of
the breakup states are consistent with a rotational band for two touching $^6$He
spheres. The steeper trajectory for a spherical $^{12}$Be is shown for comparison.}
%\mockfigure{0.4\textheight}{be12 breakup}
~\\
\begin{center}
~\psfig{file=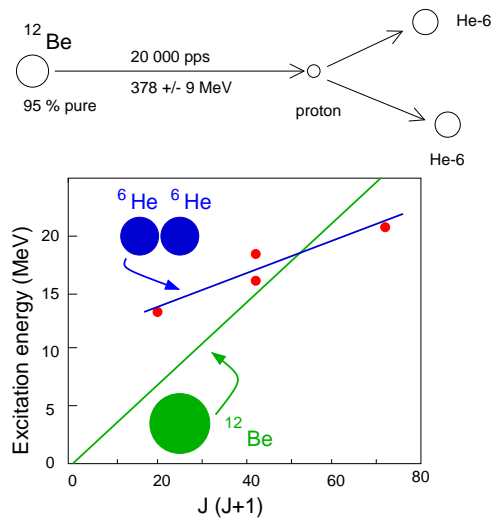,height=0.35\textheight}
\end{center}
\end{figure}

The beryllium isotopes, as discussed above, are predicted to exhibit a well
developed molecular structure in which valence neutrons bind together two
helium-like centres \cite{vonO-2,Kanada-Puri}. The molecular states are
expected near the threshold for $^4$He and $^6$He decay, consistent with the
Ikeda diagram \cite{Ikeda-Takigawa-Horiuchi}. Recently, experimental evidence
in support of these predictions has been obtained in a measurement of
p($^{12}$Be,$^{12}$Be$^*$)p using a radioactive beam \cite{be12PRL}.
A beam of 378 MeV
$^{12}$Be ions was produced at the GANIL laboratory using the LISE3
spectrometer. The fragmentation products from a primary $^{18}$O beam were
purified to produce a beam of 95\% $^{12}$Be at $2 \times 10^4$ pps. Fragments
from the binary breakup of $^{12}$Be$^*$ were identified by Z and A in
E.$\Delta $E telescopes. Measurements of the angles and energies allowed the
excitation energy of the parent $^{12}$Be to be reconstructed. Evidence was
obtained for both $^6$He$^6$He and $\alpha $$^8$He breakup, beginning at an
excitation energy just above threshold (close to 10 MeV in each case). The
$^6$He$^6$He spectrum is less complicated because the identical spin zero particles
restrict the states to even spin and parity. Unfortunately, no partial width
measurements were possible. Remarkably, the relative angular distributions of
the $^6$He$^6$He breakup fragments could be measured sufficiently accurately to
suggest spin assignments for several levels. The result was an approximate
rotational sequence (fig. \ref{be12-fig}, \cite{be12PRL}\,) reminiscent
of the results for
$^{12}$C$^{12}$C breakup of $^{24}$Mg \cite{Fulton}. The moment of inertia is
consistent with two touching $^6$He spheres and greatly exceeds that for a
spherical $^{12}$Be nucleus. Now that there is experimental evidence to support
the theoretical suggestions of clustering in the neutron rich beryllium
isotopes, it is likely that clustering and two-centre effects must be
taken into account in all studies of this region. This is relevant in view of
the recent measurements of p($^{11}$Be,$^{10}$Be)d spectroscopic factors with a
radioactive beam \cite{Lewes}, which represent the beginning of
transfer reaction spectroscopy far from stability.

\section{Outlook}

Nuclear level schemes reveal an underlying cluster structure in many
circumstances. One of the most intriguing continues to be the large-scale
structure epitomised by $^{24}$Mg as $^{12}$C+$^{12}$C. Evidence exists that
some excited states in $^{24}$Mg have a preference (beyond the statistical
probability) for decay into two $^{12}$C fragments, but the interpretations are
open to ambiguities. A key requirement is for partial decay widths such as
$\Gamma $($^{12}$C)/$\Gamma $ to be measured for breakup states. The recent
experiments using $^{12}$C($^{16}$O,$^{12}$C$^{12}$C)$\alpha $ and
$^{12}$C($^{12}$C,$^{12}$C$\gamma $)$^{12}$C give some indication of how this
might be done. Meanwhile, as experimental techniques advance, $^{12}$C+$^{12}$C
scattering continues to show remarkable behaviour such as the fully aligned
angular momentum in mutual 3$_1^-$ scattering. The
recent 3$_1^-$3$_1^-$ experiment
probed $^{24}$Mg at high spins, beyond the normal fission limit, in a regime
where less ambiguities may exist owing to the lower level density.

Probably the most exciting prospects in cluster studies are amongst the light,
very neutron rich nuclei.
The AMD calculations have finally taken clustering away from magic shells and
the N=Z line, and away from cluster models themselves. The importance of
clustering up until now has largely been in the computational power that it
provides when the clustering symmetry in the structure is recognised in the
formulation of the model. What is somewhat different about the AMD work is that
it implies, with no {\em a priori} assumptions, that it is necessary to take
clustering into account in order to get any proper understanding of the
structure of light neutron rich nuclei, even in their ground states.
The recent results for the $^6$He$^6$He breakup of $^{12}$Be point the way to
many exciting experiments in the future, using radioactive beams to explore
this region far from stability in a way that was not possible with stable
beams.

\begin{acknowledgments}
The work described here derives from the author's work over a number
of years with the CHARISSA collaboration, and their support and
contributions are gratefully acknowledged. Thanks also to Sharpey, for
encouragement back in the thesis days.
\end{acknowledgments}

{\normallatexbib

\bibliographystyle{plain}
\chapbblname{catfordbib}
\chapbibliography{logic}

}

%\newpage

\end{document}